\documentclass[useAMS,usenatbib]{mn2e}

\usepackage{graphicx}

\title[An accretion disc origin for the `XBLR' in 1H0707-495]{An accretion disc origin for the `X-ray broad line region' in 1H0707-495}
\author[A. J. Blustin \& A. C. Fabian]{A. J. Blustin$^{1}$\thanks{E-mail: ajb@ast.cam.ac.uk (AJB)} and A. C. Fabian$^{1}$ \\
$^{1}$Institute of Astronomy, University of Cambridge, Madingley Road, Cambridge CB3 0HA \\
}
\begin{document}

\date{Accepted 2009 August 22.  Received 2009 August 21; in original form 2009 August 4}

\pagerange{\pageref{firstpage}--\pageref{lastpage}} \pubyear{2009}

\maketitle

\label{firstpage}

\begin{abstract}
We use a 380~ks XMM-Newton high-resolution RGS spectrum to look for narrow spectral features from the nuclear environment of 1H0707-495. We do not find any evidence of a line-of-sight ionized wind (warm absorber). We do, however, detect broad emission lines, of width $\sim$5000~km~s$^{-1}$, consistent with O~VIII~Ly$\alpha$, N~VII~Ly$\alpha$, C~VI~Ly$\alpha$ and a Fe~XIX/Fe~XX/Ne~IX~He$\alpha$ blend. Intriguingly, these lines have both blueshifted and redshifted components, whose velocity shifts are consistent with an origin in an accretion disc at $\sim$1600~R$_{g}$ from the black hole. The features can be interpreted as the narrow line cores of the disc reflection spectrum, thus providing independent support for the discline interpretation of the X-ray spectrum of 1H0707-495. We discuss the relevance of our findings for the `X-ray broad line region' in other Seyferts, and for the origins of the optical broad line region itself.
\end{abstract}


\section{Introduction}
\label{introduction}

The narrow line Seyfert 1 galaxy 1H0707-495 (z=0.0411; \citealt{remillard1986}) has been a popular target for X-ray observations due to its enigmatic spectrum and high variability. Most attention has been focused on spectral features above 6~keV, which have been interpreted as being due to either deep partial covering absorption \citep[e.g.][]{boller2002} or relativistically-broadened Fe~k$\alpha$ line emission from the inner accretion disc \citep[e.g.][]{fabian2004}. Most recently, analysis of a large XMM-Newton dataset has supported the relativistic discline interpretation for the spectral and variability properties, and has also shown that the soft X-ray band contains a feature consistent with a discline from Fe-L, the first ever such detection \citep{fabian2009,zoghbi2009}.

In this paper, we investigate a $\sim$380~ks high-resolution soft X-ray RGS spectrum from the January-February 2008 XMM-Newton observations of 1H0707-495. RGS spectra of Seyferts frequently contain narrow line absorption from ionised winds \citep[e.g.][]{blustin2005}, narrow line emission from a kiloparsec-scale region co-spatial with the optical narrow line region \citep[e.g.][]{ogle2000}, and sometimes also broader emission lines \citep[see~e.g.][~and~references~therein]{costantini2007}. These latter features have velocity widths consistent with those of the optical broad line region (BLR), but it is as yet unclear whether the optical and X-ray lines share a common origin. We search for evidence of these various features in 1H0707-495, and discuss the implications for the nuclear region of this AGN.

\section{Data}
\label{data}

\begin{figure*}
\includegraphics[width=105mm,angle=-90]{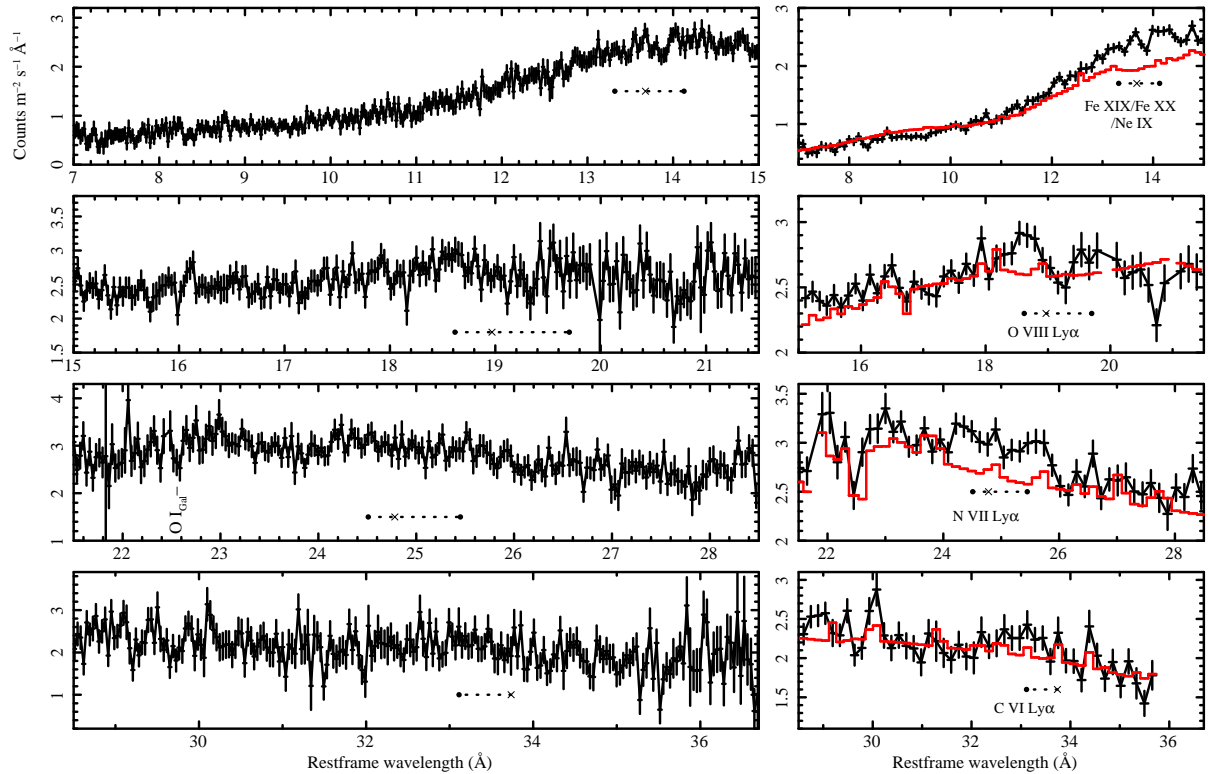}
 \caption{Fluxed restframe RGS spectrum of 1H0707-495, 380~ks exposure time, with RGS1 and RGS2, first and second spectral orders combined: left column, in bins of three (first-order) channels; right column, in bins of 12 channels. The fitted centroids of the broad emission line velocity components are marked with solid dots, and the laboratory wavelengths of the transitions with crosses. The position of the Galactic O I absorption line is also marked. The red lines in the right panels are the continuum model described in section~\ref{broad_features}.}
\label{total_spec}
\end{figure*}

We used data from four XMM-Newton observations of 1H0707-495 which took place in early 2008. The observation IDs, start dates and amounts of RGS1 good time (after background filtering) are listed in Table~\ref{obs_details}. In each case, the data were processed using {\sc sas} version 7.1.0 (xmmsas\_20070708\_1801), and the event lists were filtered to remove intervals affected by high proton background. RGS spectra were obtained using rgsproc, and these spectra, from both RGS1 and RGS2 and including first and second spectral orders, were combined using the method of \citet{page2003} into a single spectrum representing $\sim$380~ks of exposure time, and rebinned into groups of three (first-order) spectral channels. This combined spectrum is shown in Fig.~\ref{total_spec}. Either {\sc spex} version 2.00.11 \citep{kaastra1996} or {\sc xspec} version 12.5.0aj \citep{arnaud1996} were used for model generation and spectral fitting, as specified below. Spectral fitting was performed using the C-statistic \citep{cash1979}, and uncertainties are 1$\sigma$, unless otherwise stated.

\begin{table}
 \centering
 \begin{minipage}{80mm}
  \caption{\emph{XMM-Newton} observations of 1H0707-495: Observation ID; start date; RGS1 good exposure time in seconds, with the percentage of total exposure that these represent.}
\label{obs_details}
  \begin{tabular}{@{}lll@{}}
  \hline
Obs ID & Date & RGS exp (percent GT)\\
\hline
0511580101 & 2008-01-29 & 110589 (89)  \\
0511580201 & 2008-01-31 & 94202 (80)  \\
0511580301 & 2008-02-02 & 92952 (85)  \\ 
0511580401 & 2008-02-04 & 82586 (78)  \\ 
Total good time &       & 380330 (83) \\
\end{tabular}
\end{minipage}
\end{table}

\section{Analysis and results}
\label{results}

\subsection{Narrow spectral features}
\label{narrow_features}

\begin{figure}
\includegraphics[width=58mm,angle=-90]{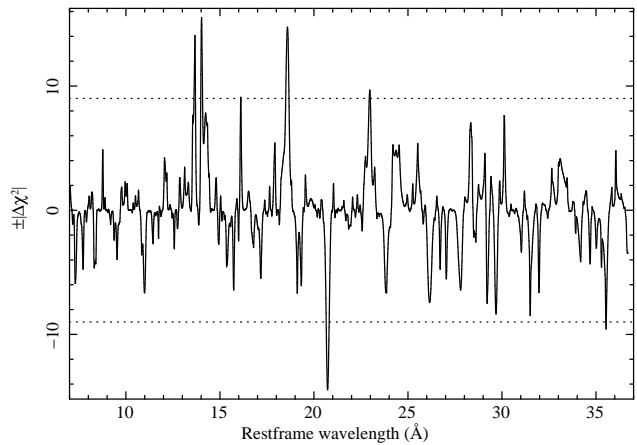}
 \caption{The statistical significance of narrow ($\leq$1000 km~s$^{-1}$) spectral features in the RGS spectrum of 1H0707-495; the $|\Delta\chi^{2}|$ refers to change in fit statistic upon addition of a narrow gaussian at the relevant wavelength. Positive values refer to emission features and negative ones to absorption features. The dotted lines indicate $\pm|\Delta\chi^{2}|=\pm9$, which is equivalent to 3$\sigma$ significance.}
\label{feature_sig}
\end{figure}

We searched for narrow ($\leq$1000 km~s$^{-1}$) spectral features in the combined RGS spectrum, using a routine (courtesy of M. J. Page) which moves across the spectrum assessing the significance of a gaussian fit against a local continuum in narrow spectral ranges (Fig.~\ref{feature_sig}). It found seven features significant at $\geq3\sigma$, and a further 33 features significant at between $2-3\sigma$. This is broadly consistent with what would be expected by chance in an RGS spectrum of $\sim$1000 bins, although the seven features at $\geq3\sigma$ are more than the approximately three which would be expected. If the apparent emission features are real, they could be consistent with Ne~IX and various states of L-shell iron, although we cannot find a consistent interpretation for them all. There is no plausible identification for the possible absorption line at $\sim$20.7~\AA. In general, the soft X-ray spectrum contains no evidence of a warm absorber type outflow.

\subsection{Broad spectral features}
\label{broad_features}

\begin{figure}
\includegraphics[width=60mm,angle=-90]{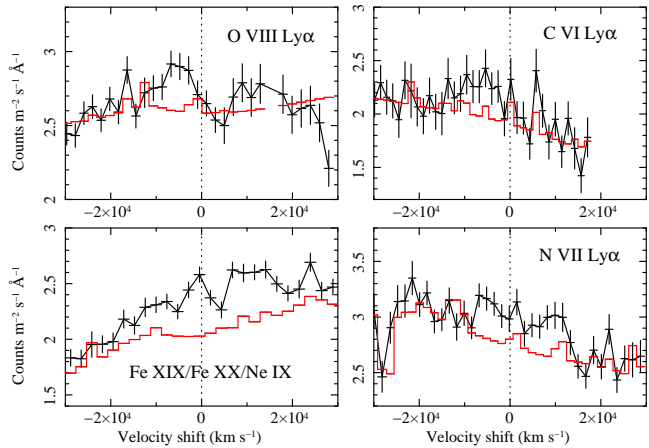}
 \caption{Plots in velocity space of broad emission features in the spectral regions of O~VIII~Ly$\alpha$, N~VII~Ly$\alpha$, C~VI~Ly$\alpha$ and Fe~XIX/Fe~XX/Ne~IX~He$\alpha$. Black points are data, and the red line in each case is the continuum model described in section~\ref{broad_features}, renormalised to local line-free regions as necessary. Zero velocity corresponds to the laboratory wavelength of each transition, except in the case of the Fe~XIX/Fe~XX/Ne~IX~He$\alpha$ blend which is taken to have a rest frame wavelength of 13.677~\AA\ (see section~\ref{broad_features}). Blueshifts are negative.}
\label{broad_features_plot}
\end{figure}

In a spectrum rebinned by a factor 12 in terms of original channels (i.e. by a factor four from the standard three times rebinning), broad emission features are clearly visible close to the wavelengths of transitions typically observed in soft X-ray spectra of Seyferts. Fig.~\ref{broad_features_plot} shows velocity space plots of the spectral regions of O~VIII~Ly$\alpha$, N~VII~Ly$\alpha$, C~VI~Ly$\alpha$ and an Fe-L/Ne~IX~He$\alpha$ line blend. We identify this latter feature as principally a combination of Fe~XIX (13.54~\AA), Fe~XX (13.79~\AA) and the Ne~IX He-like triplet lines, and take its average wavelength to be 13.677~\AA, which is its peak wavelength predicted by the reflection models described in section~\ref{xblr_properties}. 

Possible emission features are observed both bluewards and redwards of the rest wavelengths, in all cases except C~VI~Ly$\alpha$. The clearest example of a double-peaked line is N~VII~Ly$\alpha$. Any red wing of C~VI~Ly$\alpha$ would be in a range where the RGS effective area is tailing off, which could be why it is not observed. In the plot of Fe~XIX/Fe~XX/Ne~IX~He$\alpha$, the feature at zero velocity is probably a blend of the narrow emission lines at $\sim$13.55 and $\sim$13.68~\AA, which can be seen in the top panel of Fig.~\ref{total_spec}; these lines may be, respectively, the intercombination and forbidden lines of the Ne~IX He-like triplet. Wavelengths, velocity shifts, velocity widths and fluxes obtained from gaussian fits to the broad features are listed in Table~\ref{broad_line_properties}. At the implied ionisation level of the emitter, we would also expect to observe O~VII~He$\alpha$. There is no clear detection of this in our spectrum, since it falls in a region with only one operational CCD, and is confused by the presence of the Galactic O edge. We list upper limits to its flux in Table~\ref{broad_line_properties}.

\begin{table*}
 \centering
 \begin{minipage}{175mm}
  \caption{Properties of broad emission lines in the RGS spectrum of 1H0707-495, obtained from gaussian fits to these features. Line or line blend; line component; $\lambda_{lab}$, laboratory wavelength in \AA; $\lambda_{rest}$, fitted rest-frame wavelength in \AA; $v_{shift}$, velocity shift in km~s$^{-1}$ (blueshifts are negative); $v_{FWHM}$, line FWHM in km~s$^{-1}$; $F_{-14}$, line flux in 10$^{-14}$ erg~cm$^{-2}$~s$^{-1}$. The O~VII He-like triplet blend is not clearly detected; we list upper bounds to the flux of a gaussian fitted at the expected wavelengths of its blue and red components. For this feature, we take the rest wavelength predicted by the reflionx model (see section~\ref{xblr_properties}), fixing the width at 5000~km~s$^{-1}$ and the velocity shifts at the average value of $\pm$7400~km~s$^{-1}$.}
\label{broad_line_properties}
  \begin{tabular}{@{}lllllll@{}}
  \hline
Line or line blend & Component & $\lambda_{lab}$ & $\lambda_{rest}$ & $v_{shift}$ & $v_{FWHM}$ & $F_{-14}$ \\
  \hline
O~VIII~Ly$\alpha$        & blue & 18.969 & 18.62$\pm$0.07              & -5500$^{+1000}_{-1100}$ & 4900$^{+2800}_{-2200}$    & 1.4$\pm$0.4 \\
O~VIII~Ly$\alpha$        & red  & 18.969 & 19.7$\pm$0.2                & 11600$\pm$3100        & 9100$^{+11300}_{-5600}$  & 1.6$\pm$0.7 \\
N~VII~Ly$\alpha$         & blue & 24.781 & 24.5$^{+0.2}_{-0.1}$          & -3300$^{+2100}_{-1200}$ & 9200$^{+5000}_{-2600}$    & 3.1$\pm$0.5 \\
N~VII~Ly$\alpha$         & red  & 24.781 & 25.5$\pm$0.1                & 8100$^{+1200}_{-1600}$  & 8200$^{+4400}_{-2300}$    & 2.5$\pm$0.5 \\
C~VI~Ly$\alpha$          & blue & 33.736 & 33.1$^{+0.3}_{-0.2}$          & -5500$^{+2600}_{-1700}$ & 9400$^{+7700}_{-3900}$   & 2.1$\pm$0.6 \\
Fe~XIX/Fe~XX/Ne~IX~He$\alpha$  & blue & 13.677 & 13.3$^{+0.2}_{-0.1}$    & -8100$^{+3900}_{-3200}$ & 13900$^{+13100}_{-6700}$ & 1.6$\pm$0.6 \\
Fe~XIX/Fe~XX/Ne~IX~He$\alpha$  & red  & 13.677 & 14.13$^{+0.06}_{-0.07}$ & 9900$^{+1300}_{-1500}$  & 11900$^{+4900}_{-3200}$  & 3.3$\pm$0.5 \\
O~VII He-like triplet    & blue & 22.000 & 21.591  & -7400 & 5000   & $\leq$2.0 \\
O~VII He-like triplet    & red  & 22.000 & 22.409  & 7400  & 5000   & $\leq$1.1 \\
  \hline
\end{tabular}
\end{minipage}
\end{table*}

The spectral continuum in these plots and fits is the \citet{fabian2009} relativistic discline model, with the same parameters except that the blackbody temperature and normalisation, and the powerlaw and reflionx normalisations, were re-fitted in {\sc xspec} to continuum regions of the RGS spectrum to account for the systematic calibration differences between EPIC and RGS. The resulting normalisations are respectively $N_{BB}=1\times10^{-4}$, $N_{pow}=2\times10^{-3}$ and $N_{refl}=2\times10^{-5}$ in {\sc xspec} units, and the blackbody temperature is 0.05~keV. The model was converted to a table model in {\sc spex} for convenient fitting to RGS data, and its overall normalisation was allowed to vary during the gaussian fits described above in order to obtain the best local fit to each line. The depth of the z=0 O~I line in the RGS spectrum is consistent with neutral absorption due to our Galaxy, in line with the findings of \citet{zoghbi2009}, so we set the z=0 neutral absorption to the Galactic value\footnote{Weighted average value from the Leiden/Argentine/Bonn survey of Galactic HI \citep{kalberla2005}} of 4.31$\times10^{20}$~cm$^{-2}$. 

\subsubsection{Properties of the broad line emitting gas}
\label{xblr_properties}

We can estimate the ionisation levels and elemental abundances of the broad-line emitting plasma through comparison with model spectra. In the following analysis, we fitted models to the blue and red components separately. We used the \citet{ross2005} reflionx model in {\sc xspec} to generate reflected emission-line spectra over a wide range of ionisation parameters, with the Fe/O relative abundance ratio set to 1. We chose this model, which is valid for an optically-thick reflector, because of the dynamical evidence of origin in an accretion disc (see section~\ref{discussion}). The resulting spectra were then imported into {\sc spex} (again for convenience of fitting to RGS data) as table models, and convolved with gaussian velocity broadening with $\sigma=2100$~km~s$^{-1}$. This value was chosen to match the best-constrained fitted FWHM, to the blue component of O~VIII~Ly$\alpha$, of $\sim$5000~km~s$^{-1}$. These emission line models were then added to the continuum described in section~\ref{broad_features}.

The ratio of O~VIII to O~VII emission signals the ionisation level of the emitter. Although the O~VII spectral region is ambiguous, it can nevertheless be used to constrain relative flux. We therefore fitted the models in narrow spectral regions around the wavelengths of O~VIII~Ly$\alpha$ and the O~VII H-like triplet, fitting the two sets of lines for each velocity phase separately. 

We found that the blueshifted and redshifted components require different ionisation parameters: $\xi_{blue}=200^{+660}_{-120}$ and $\xi_{red}=29^{+34}_{-27}$ erg~cm~s$^{-1}$ respectively, where $\xi=L_{\rm ion}/(nr^{2})$; $L_{\rm ion}$ is the 1-1000~Ryd (13.6~eV-13.6~keV) ionizing luminosity, $n$ is the gas density and $r$ is the distance from the ionizing source \citep{tarter1969}. This is necessarily a tentative result, since the fluxes of the O~VII lines cannot be measured directly. Also, the fitted O~VIII/O~VII ratio for the red component will be affected by what we assume for the depth of intrinsic neutral absorption, since the O~I edge falls between the expected wavelengths of the two emission lines for this component. We note, though, that no O~I absorption line is detected at the relevant wavelength in the restframe of 1H0707-495, so our assumption of no intrinsic neutral absorption is reasonable.

Having estimated the ionisation parameter of each velocity component, we can then potentially use the models at the respective ionisation levels to make rough estimates of elemental abundances with respect to solar. This is only possible for elements with unblended lines, and so we cannot reliably estimate Fe/O, or indeed N/O since N~VII~Ly$\alpha$ is blended with the C~VI radiative recombination continuum (RRC) at the resolution of our model. Only the blue component has detected lines from both oxygen and carbon. We therefore separately fitted the model normalisation in narrow spectral regions around O~VIII~Ly$\alpha$ and C~VI~Ly$\alpha$ for this component; the ratio of these normalisations gives an estimate of the elemental abundance ratio. We obtain a C/O ratio of $\sim4\pm2$ relative to solar \citep{morrison1983}.

\section{Discussion and conclusions}
\label{discussion}

\begin{figure}
\includegraphics[width=55mm,angle=-90]{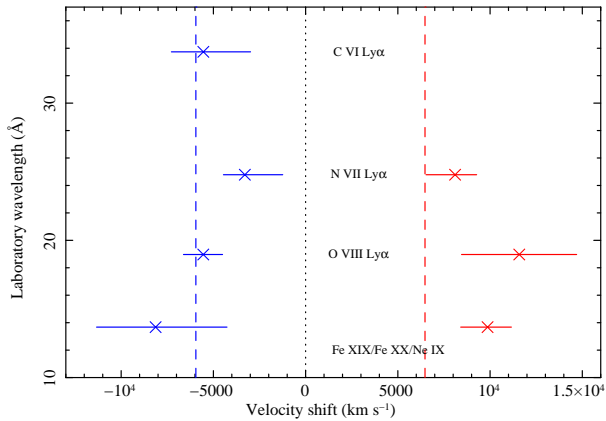}
 \caption{Velocity shifts of the blue and red components of the broad emission lines. The long-dashed lines mark the predicted velocity shifts for lines originating in an accretion disc with an inclination of 58.5$^\circ$, at 1500$-$1700~R$_{g}$ from the black hole.}
\label{vel_shifts}
\end{figure}

\begin{enumerate}
  \item The RGS spectrum of 1H0707-495 contains no evidence of narrow absorption features from a line-of-sight absorbing wind. This is unsurprising, since there is no intrinsic ionized UV absorber \citep{dunn2007}; Seyferts with a soft X-ray absorbing outflow also show evidence of an ionized wind in the UV \citep{crenshaw1999}. 

 \item There are broad (FWHM $\sim$5000~km~s$^{-1}$) emission features in the spectrum, which correspond to O~VIII~Ly$\alpha$, N~VII~Ly$\alpha$, C~VI~Ly$\alpha$ and a blend of Fe~XIX/Fe~XX/Ne~IX~He$\alpha$. The features have both blueshifted and redshifted components, with an average velocity shift of magnitude $\sim$7400~km~s$^{-1}$. With this line-of-sight velocity, and assuming that the emitting region has the same inclination to our line of sight as the inner accretion disc ($i$=58.5$^\circ$ as obtained from fits to EPIC data; \citealt{zoghbi2009}), we can derive the actual orbital velocity of the emitter, and then use the virial theorem to calculate a distance of $\sim6\times10^{14}$~cm from the black hole. This is equivalent to $\sim$1600 gravitational radii (R$_g$), for a black hole mass of $10^{6.37}$~M$_{\odot}$ \citep{zhou2005}. This distance, plus the existence of both blueshifted and redshifted components, show that the lines must originate in the accretion disc. The double-peaked lines are produced by doppler-shifted emission from material moving towards and away from us on respective sides of the rotating disc. For a disc around a black hole, relativistic effects also come into play: the flux of the blue wing of the line is increased by doppler boosting, and gravitational redshift and time dilation move the lines to longer wavelengths, creating asymmetric velocity shifts \citep[see~e.g.][]{fabian2000}. These effects become less important with increasing distance from the black hole, so that they will not be highly significant for lines originating at $\sim$1600~R$_g$. We calculated the expected velocity shifts of the lines using the diskline model in {\sc xspec}, assuming the lines originate in a region at 1500$-$1700~R$_g$; these are overplotted on the measured velocity shifts in Fig.~\ref{vel_shifts}. Some of the variability in velocity shift between lines is likely to be due to line blending within the broad features, especially in the case of N~VII~Ly$\alpha$, which is blended with the C~VI RRC.

 \item The emission features in the RGS could be the narrowest parts of the line cores, originating far out in the disc, of the broadband soft X-ray reflection spectrum identified in EPIC data. A similar interpretation has been advanced in the past for some Fe k$\alpha$ emission line components observed in certain Seyferts \citep[e.g.][]{nandra2006}. Our findings thus provide independent support for the conclusions of \citet{fabian2009} and \citet{zoghbi2009} regarding the discline interpretation of the X-ray spectrum. 

  \item The red and blue wings of the emission lines appear to originate in gas at different ionization levels. If this is a real effect, it implies density variations in the disc, and/or differences in illumination, perhaps as a result of disc warping or a patchy corona. The blending of lines within the relevant broad emission features prevents us from coming to any conclusions about the Fe/O or N/O abundance ratios in the emitter, or making any direct comparison with the $\sim$9 times overabundance of iron in the inner disc, as reported by \citet{zoghbi2009}. In the case of carbon, though, there is some indication of a supersolar abundance of $\sim$4$\pm$2 with respect to oxygen.

  \item Soft X-ray emission lines with velocity widths of a few thousand km~s$^{-1}$ have been detected in several Seyfert galaxies, including NGC~4051 \citep{ogle2004}, NGC~5548 \citep{steenbrugge2005}, Mkn~509 \citep{smith2007}, Mkn~279 \citep{costantini2007}, and Mkn~335 \citep{longinotti2008}. They are often interpreted as originating in the BLR. The likelihood that the soft X-ray broad lines in 1H0707-495 are produced in the accretion disc may not be relevant to these other sources since none of their broad lines are double-peaked. On the other hand, the 1H0707-495 broad lines may simply originate at a smaller disc radius, due to the small black hole mass in this source. 

  \item What is the wider relevance of soft X-ray line reflection from the outer disc? Modelling of high-resolution radio observations of the nearby prototypical Seyfert 2 NGC~1068 \citep{gallimore2004} has shown that the radio emission from the inner 0.4~pc is likely to originate in an X-ray irradiated molecular disc, where gas is being heated and a radio bremsstrahlung-emitting disc wind is being driven off. Their models predict that soft X-ray emission lines should be observable. This region in NGC~1068 is far larger than that from which the 1H0707-495 emission lines originate. Nevertheless, our connection of broad soft X-ray emission lines with the accretion disc makes it tempting to speculate that the Seyfert `X-ray BLR' could be related to the outer disc X-ray emission predicted by \citet{gallimore2004}. We also note that the blueshifted UV C~IV emission ($v_{shift}\sim$-2000~km~s$^{-1}$, $v_{FWHM}\sim$5000~km~s$^{-1}$) in 1H0707-495 has been interpreted as originating near the base of an accretion disc wind \citep{leighlymoore2004,leighly2004}. The UV lines lack a redshifted component, implying that they are produced further out (or higher above the disc) where wind emission dominates disc emission. The lower blueshift of the UV lines implies that the wind is decelerating with distance, or flowing at a greater inclination to our line of sight. The XBLR would then be the link between the accretion disc and the base of a disc wind, which is perhaps the ultimate source of the optical broad line region itself. 
\end{enumerate}

\section*{Acknowledgments}

AJB and ACF acknowledge the support of, respectively, an STFC Postdoctoral Fellowship and the Royal Society. AJB thanks Randy Ross for providing information about spectral features in the reflionx model. We thank the referee, Katrien Steenbrugge, for useful comments.

\label{lastpage}

\end{document}